\documentclass[%
 reprint,
 amsmath,amssymb,
 aps,
]{revtex4-2}

\usepackage[utf8]{inputenc}
\usepackage{graphicx}% Include figure files
\usepackage{dcolumn}% Align table columns on decimal point
\usepackage{bm}% bold math
\usepackage[colorlinks=true,citecolor=red,filecolor=green,linkcolor=blue]{hyperref}
\usepackage{xcolor}

\begin{document}

\preprint{APS/123-QED}

\title{Stochastic dengue outbreak modeling and study using nonlinear analytical techniques}

\author{
Tongbram Surkumar Singh$^{1}$,
Mairembam Kelvin Singh$^{1}$,
R.K. Brojen Singh$^{2}$
Moirangthem Shubhakanta Singh$^{1,*}$%
}

\affiliation{$^{1}$Department of Physics, Manipur University, Canchipur - 795003, Imphal, Manipur, India.\\
$^2$School of Computational and Integrative Sciences, Jawaharlal Nehru University, New Delhi-110067, India.}

\email{mshubhakanta@yahoo.com}

\begin{abstract}
{\noindent}One of the present challenges of dengue outbreak is the real data driven modeling of the disease transmission dynamics which can capture the complex dynamical characteristics for accurate prediction and control. However, the conventional validation of such model is primarily based on statistical agreement between model simulated and reported incidence trajectories. In this work, we propose a seasonally driven stochastic dengue transmission model which is the modified stochastic version of the deterministic SEIR-SEI dengue outbreak model. The simulation of the proposed model is done using classical stochastic simulation algorithm. Then, the model validation is done by comparing these simulated data with the reported laboratory confirmed dengue incidence data at Imphal West district, Manipur for the three consecutive years (2023-2025) by systematic analysis using various nonlinear dynamical validation techniques. The model's performance is evaluated using coefficient of determination ($R^2$) and relative error ($RE$), followed by sensitivity analysis to assess parameter robustness. The calibrated model successfully reproduces the reported outbreak trajectories for all the three years, achieving coefficients of determination greater than 0.96 and relative error less than $7\%$. The sensitivity analysis reveals that the transmission activation time is the most important transmission modulation parameter influencing calibration performance. Both the complexity-entropy and recurrence analyses show that the calibrated stochastic model captures substantial features of the temporal organization of the reported outbreak while retaining measurable significant differences in higher-order dynamical measures and providing complimentary information beyond conventional goodness-of-fit measures and offer a more comprehensive framework for validating stochastic epidemic models.\\


\end{abstract}

\maketitle

\section{Introduction}
{\noindent}The dynamics of infectious diseases during an epidemic or pandemic is a complex process in two broad ways: first, how the pathogen attacks our body through a cascade of pathways that reach specific gene/s \cite{mogensen2009pathogen}, and second, how the disease spreads in the network of individuals in a certain demographic region or throughout the world \cite{pastor2015epidemic}. One set of such types of diseases is vector-borne infectious diseases which remain as an major public health concern, particularly in tropical and subtropical regions where climatic conditions favor disease transmission \cite{who_vector,gubler1998dengue,WHO_vector_2024}. Among these infectious diseases, dengue is considered one of the fastest growing and highly cantagious arboviral infections worldwide \cite{bhatt2013,WHO_dengue_2023}. This disease is caused by the infection with dengue virus which is a single-stranded, positive-sense RNA virus belonging to the genus Flavivirus of the family Flaviviridae \cite{schaefer2024dengue}. Dengue virus has four genetically related but antigenically distinct serotypes, namely, DENV-1, DENV-2, DENV-3, and DENV-4, each capable of causing the full spectrum of clinical disease \cite{paz2024dengue}. Generally, Dengue virus attacks and destroys the host's innate immune defense network, that is the interferon signaling pathway, and by interacting with the key host genes related to pathogen sensing and inflammation \cite{kao2018dengue}. Further, it has been reported that the dengue serotypes, DENV-2 and DENV-3 cause severe dengue manifestations, such as, plasma leakage, hemorrhage fever, organ dysfunction, shock syndrome etc \cite{schaefer2024dengue}.\\

{\noindent}The dengue transmission from one individual host to another is primarily due to the bite by the infected female mosquitoes, \textit{Aedes aegypti} and \textit{Aedes albopictus}, which thrive in urban and semi-urban environments~\cite{gubler1998dengue,Kraemer2015,bhatt2013}. Rapid urbanization, lack of sanitation, increased human mobility, population growth, uncontrolled growth of mosquitoes, and climate variability have collectively drive the global risk of dengue epidemic/pandemic, resulting in recurrent seasonal outbreaks that continue to challenge public health systems~\cite{Messina2019,WHO_dengue_global_2025,stanaway2016}.\\

{\noindent}Dengue disease outbreak emerges from complex interactions among human hosts and mosquito vectors facilitated by the environmental conditions, lack of precautionary measures to control the mosquito vectors and seasonal climate variability~\cite{bhatt2013,Messina2019}. The pattern of these interactions generates a rapid disease spreading complex epidemic network, where, each infected individual has nonlinear character with uncontrolled increase in case incidence during epidemic \cite{ni2024epidemiological}. However, once proper control measures are taken up carefully, the dengue epidemic is followed by an endemic with epidemic decline~\cite{stanaway2016}. This complex dynamics of the dengue epidemic can be modeled using various methods, namely, deterministic approach \cite{andraud2012dynamic,Scheffer2009,Strogatz2015}, stochastic techniques \cite{de2013stochastic}, hybrid model \cite{garzon2025spatio}, game theoretical approach \cite{sun2025modeling}, fractional calculus \cite{pooseh2011fractional}, reaction-diffusion method \cite{wang2011nonlocal} etc. These methods are proposed in order to understand the mechanism of dengue outbreak, its dynamical behavior, forecasting and then to control the epidemic/pandemic with special concern on public health, economy and various preventive measures.\\

{\noindent}Data driven mathematical models have been quite successful to understand the dynamical behavior of the dengue epidemic leading to the prediction and control of the epidemic and hence, they are central to infectious disease research~\cite{Anderson1991}. In dengue epidemiology in particular, the $SEIR-SEI$ (susceptible exposed infected recovered-susceptible exposed infected) framework is the model which is widely used to investigate dengue transmission dynamics, epidemic thresholds, and possible intervention strategies~\cite{Hethcote2000,Smith2012,vandenDriessche2002}. While this deterministic model describes the classical mean-field epidemic behavior, the dynamics of the real dengue outbreaks are inherently stochastic in nature because the transmission due to individual level interaction events in the epidemic network occur in probabilistic manner \cite{Allen2008}. Consequently, stochastic formulations, including reaction-network models, which can be simulated using Gillespie's stochastic simulation algorithm (SSA), provide a more realistic quantitative representation of outbreak variability and transmission uncertainty \cite{Nasell2002,Gillespie1977,Otero2010,DengueBayesian2022,DengueSpatial2022}.\\

{\noindent}Next, the stochastic epidemic models are typically validated by mapping the model equations to the large scale real dengue epidemic data, and then using statistical analysis one can characterize the correctness of the model framework \cite{mayer1993statistical}. Although the statistical metrics assess agreement between simulated and observed case counts (data), they provide limited insight into whether the simulations of the model framework reproduce the underlying temporal behavior of the real epidemic dynamics. Hence, even though different dynamical realizations can generate disease incidence curves with similar statistical properties, the conventional calibration could be insufficient for validating the real dynamical patterns of the epidemic model frameworks \cite{shaman2013,viboud2018}.\\

{\noindent}Now, the complexity in the multi-dimensional data can be analyzed using statistical complexity analysis technique to characterize both real epidemic data as well as model generated data \cite{lopez1995statistical}. This probabilistic analysis of complexity of the nonlinear and nonstationary time series provides the statistical description of the dynamics at any given scale which can reveal the accessible states and probability distribution at that scale \cite{rosso2007distinguishing}. Following this technique, the complexity-entropy analysis generally quantifies the balance between randomness and temporal organization through permutation entropy and statistical complexity \cite{bandt2002permutation,rosso2007,zanin2012}, while recurrence quantification analysis (RQA) measures determinism, persistence, and predictability from phase-space trajectories~\cite{marwan2007}. Together, these methods enable the evaluation of epidemic dynamics beyond statistical agreement by characterizing their intrinsic temporal organization.\\

{\noindent}Despite the growing use of stochastic epidemic models and nonlinear dynamical analysis on the real epidemic data, there is a gap of systematic validation of the model on the data with significant accuracy. Further, only few studies have integrated the approaches mentioned above to examine whether calibrated epidemic models reproduce the nonlinear dynamical characteristics observed in the empirical outbreak data \cite{small2005}. To address these gaps is quite important to ensure the correctness of the model with the real dengue epidemic data. Then, the further analysis of the model could provide proper dengue epidemic forecast and systematic control mechanism of the dengue epidemic forcing to endemic.\\

{\noindent}In this study, we address this gap by proposing an equivalent stochastic epidemic model of the nonlinear dynamical dengue epidemic model. We use laboratory-confirmed dengue incidence data from Imphal West district, Manipur, India, during the 2023-2025 dengue outbreak season for our study. We, then, developed a stochastic framework of the SEIR-SEI dengue epidemic model formulated as a reaction network and simulated with Gillespie algorithm (SSA) using a set of parameter values assigned to the reaction network. To check the correctness of the proposed model, the simulation results of the proposed stochastic model is calibrated independently for each outbreak, using time-dependent transmission rates that account for seasonal modulation. Further, we use complexity-entropy and RQA to assess the dynamical consistency of reported and simulated outbreak data. Our results demonstrate that combining stochastic epidemic modeling with proper use of the parameter values provides a more comprehensive assessment of dengue epidemic dynamics and properties which can be for epidemic forecast and control strategies.\\

\begin{figure*}
    \centering
    \includegraphics[scale=0.8]{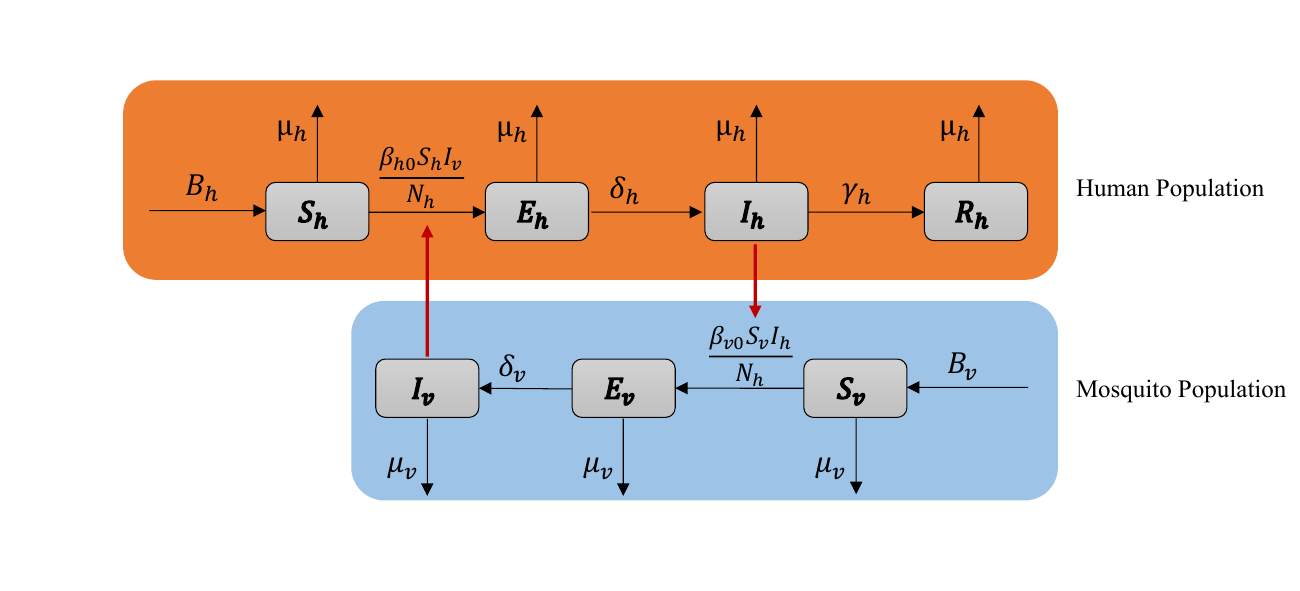}
    \caption{A schematic diagram to illustrate the SEIR--SEI compartmental structures and interaction between different compartments in the dengue transmission dynamics.}
    \label{fig1}
\end{figure*}

{\noindent}The paper is organized as follows. Section~\ref{sec:meth} presents the methodology used in this work which includes the description of the dengue incidence data, the deterministic SEIR-SEI transmission model, corresponding proposed stochastic reaction-network formulation, and the calibration process. Section~\ref{sec:results} presents the calibration of the results, sensitivity analysis, and the nonlinear characterization of the reported and simulated outbreaks. The implications of the findings are discussed in Section~\ref{sec:discussion}. Finally, the main conclusions of the study are summarized in Section~\ref{sec:conc}.

\section{Method}
\label{sec:meth}
{\noindent}We describe various methods which we employed to understand dynamics of the dengue spreading in Imphal, Manipur including data collection and modeling, mathematical model to describe the modeled data for possible control the epidemic and prediction. \\

\subsection{Data requisition and analysis}
\label{data}
{\noindent}We collected the registered dengue cases empirical data from the State Malaria Office, the National Vector Borne Disease Control Programme (NVBDCP) under the Directorate of Health Services, Government of Manipur, India~\cite{NCVBDC_Manipur_2026} for the present study. The dataset we prepared for our analysis comprises of daily laboratory-confirmed dengue cases reported from the Imphal West district during the three consecutive outbreak years. Imphal West district was chosen because continuous dengue surveillance records were available throughout the study period, and the district had experienced recurrent dengue outbreaks in recent years. We use the daily dengue-confirmed incidence cases as the function of time as the primary observable time series because it reflects the temporal evolution of dengue transmission and serves as the basis for the analysis in the present study. We also examined the incidence records independently for each outbreak year and combine to create a continuous time series covering the entire 2023-2025 period in order to examine both annual and multi-year outbreak dynamical properties. We also ensure missing data points in the prepared dataset with available incidence records to avoid any temporal inconsistencies. The complete dataset contains 1096 daily observations of dengue incidence cases covering the entire three consecutive dengue transmission seasons. \\

{\noindent}The prepared dengue incidence dataset is used for analysis using various time series analytical techniques, such as, complexity-entropy and recurrence-based analyses presented in this work. These techniques can able to explore various properties, patterns associated with the time series data.

\subsection{Deterministic SEIR--SEI Model}
{\noindent}Dengue transmission is due to the bite of mosquitoes from one individual human to another, and hence this transmission dynamics can be well described by the human-vector interaction model that captures the biological interactions between humans and vectors. Therefore, we consider a compartmental susceptible-exposed-infected-recovered-susceptible-exposed-infected (SEIR-SEI) model \cite{OTERO201187,ERICKSON20102899} in which the total human population ($N_h$) is divided into susceptible ($S_h$), exposed ($E_h$), infectious ($I_h$) and recovered ($R_h$) classes, while the total mosquito population ($N_v$) is divided into susceptible ($S_v$), exposed ($E_v$) and infectious ($I_v$) classes. This model assumes that the infection is spread through mosquito bites, with no direct human-to-human transmissions~\cite{gubler1998dengue}. After infection, there are main two incubation periods, first, extrinsic incubation period in mosquitoes during which the virus taints the cells to replicate to the mosquitoes tissues (period is generally 8-10 days), and second, intrinsic incubation period in human which is the time for onset of illness (period is approximately 7-8 days) \cite{guzman2016dengue,esteva1998analysis}. Infected humans recover and develop immunity over the duration of the outbreak for that certain serotype (but still susceptible to the other serotypes), whereas, mosquitoes have a short lifespan, and so their recovery is not considered in the model  \cite{esteva1998analysis}. The model is a compartmental model in which both human and mosquito populations are subject to constant per capita recruitment and natural death rates, where, the interaction among the compartments causes dengue transmission and related activities \cite{esteva1998analysis}. The model also assumes that the total human and mosquito populations remain constant in due course of time \cite{esteva1998analysis}, and, hence, $N_h = S_h+E_h+I_h+R_h$ and $N_v = S_v+E_v+I_v$, due to a balance between recruitment and natural mortality. A schematic diagram illustrating the compartmental structure and transitions of the interaction in the model is presented in Figure~\ref{fig1}.\\

{\noindent}The governing equations of the model represented by the Figure~\ref{fig1} are given by,
\begin{eqnarray}
\label{deterministic}
\left.
\begin{aligned}
    \frac{dS_h}{dt} &= B_h - \beta_{h0}\frac{S_h I_v}{N_h} - \mu_h S_h, \\
    \frac{dE_h}{dt} &= \beta_{h0} \frac{S_h I_v}{N_h} - (\delta_h + \mu_h) E_h, \\
    \frac{dI_h}{dt} &= \delta_h E_h - (\gamma_h + \mu_h) I_h, \\
    \frac{dR_h}{dt} &= \gamma_h I_h - \mu_h R_h, \\
    \frac{dS_v}{dt} &= B_v - \beta_{v0 }\frac{S_v I_h}{N_h} - \mu_v S_v, \\
    \frac{dE_v}{dt} &= \beta_{v0} \frac{S_v I_h}{N_h} - (\delta_v + \mu_v) E_v, \\
    \frac{dI_v}{dt} &= \delta_v E_v - \mu_v I_v.
\end{aligned}
\right\rbrace
\end{eqnarray}
with the non-negative initial conditions, $S_h, E_h, I_h, R_h, S_v, E_v, I_v \geq 0$. In this dengue transmission model described in model equations \eqref{deterministic}, the nonlinear infection terms $\displaystyle\beta_{h0} \frac{S_h I_v}{N_h}$ and $\displaystyle\beta_{v0} \frac{S_v I_h}{N_h}$ represent the transmission of infections between the two populations. $\beta_{h0}$ and $\beta_{v0}$ denote the transmission rates from infectious mosquitoes to humans and from infectious humans to mosquitoes, respectively. These two parameters are the products of the corresponding average daily biting rate and the transmission probability from the host to the vector or vice-versa \cite{Andraud2012}. The parameters $\delta_h$ and $\delta_v$ correspond to the progression rates from the exposed to infectious populations, and $\gamma_h$ represents the recovery rate of infected humans. The natural death rates of humans and mosquitoes are denoted by $\mu_h$ and $\mu_v$ respectively, and the recruitment rates are given by $B_h = \lambda_h N_h$ and $B_v = \lambda_v N_v$, with $\lambda_h$ and $\lambda_v$, the per capita birth rates of humans and mosquitoes respectively. The parameter values used in the model are summarized in Table~\ref{Table1}.\\

\subsection{Stochastic formulation of the SEIR-SEI dengue model}
{\noindent}The deterministic SEIR-SEI model represented by the equations \eqref{deterministic} is the mean-field model that provides qualitative behavior of the dynamics of dengue transmission. In order to study the quantitative behavior of the dynamics, where, the significant role of the random fluctuations are incorporated, we formulate the dengue transmission process depicted by the equations \eqref{deterministic} into a set of elementary reactions representing a network. In this representation, epidemiological transitions are treated as discrete random events that occur with specific probabilities. Each reaction event in the reaction network allows to change the system's state by modifying one or more compartment populations which involve in the reaction event. The corresponding reaction network of the deterministic SEIR-SEI model comprises of fourteen reaction channels that describe recruitment, infection, disease progression, recovery, and natural mortality in both human and mosquito populations respectively. The reactions are given in the Table \ref{Table1}. 

\subsection{Stochastic simulation algorithm to simulate reaction network}
{\noindent}The dynamics of the reaction network given in Table \ref{Table1} can be studied using the multi-variable Master equation representing the reaction network which is generally very difficult to solve analytically except for simple systems \cite{gardiner1985handbook}. However, such reaction networks either constructed from the Master equation or deterministic systems can be numerically solved using standard stochastic simulation algorithm (SSA) developed by Gillespie \cite{Gillespie1977}. Since the theoretical foundation of this algorithm was developed by Doob \cite{doob1942topics,doob1945markoff} and the algorithm was originally introduced by Kendall \cite{kendall1950artificial}, this SSA is generally known as Doob-Gillespie algorithm or widely Gillespie algorithm. Gillespie systematically formulated this algorithm in detail beautifully \cite{Gillespie1977} as workable simulation method and we are explaining briefly the algorithm as follows. Consider a system of size $V$ having $N$ molecular species interacting through a well defined set of $M$ reaction channels. If the state vector of the system is defined by a molecular species population vector $\vec{X}=\left[x_1,x_2,\dots,x_N\right]^T$, where, $T$ is the transpose of the vector, then, the dynamics of the system is governed by the time evolution of this state vector induced by the birth and death process of the molecular species in due course of firing of the reactions \cite{Gillespie1977,gillespie1976general}. This time evolution of the state vector describing the system can be systematically simulate using SSA. SSA is a Monte carlo like algorithm designed based on two fundamental random processes inherent to any system of randomly interacting molecular species: random firing of a reaction ($\mu$) and random reaction time ($\tau$) which are statistically independent of each other \cite{Gillespie1977}. In the algorithm these two random processes are realised by generating two sets of statistically independent, uniform random numbers $r_1$ and $r_2$ respectively. Then, the joint probability of the system can be defined by $P(\tau,\mu)=P(\tau)P(\mu)$. Putting this expression to the Master equation of the system, one can obtain the reaction time $\tau$ as $\tau=\frac{1}{a_o}ln\left[\frac{1}{r_1}\right]$, where, $a_o=\displaystyle\sum_{i=1}^{M} a_i$, with $a_i$ as the propensity function of the ith reaction channel. This $a_i$ can be related to stochastic rate constant $c_i$ and possible number of molecular combinations $h_i$ by $a_i=h_ic_i$. Moreover, this stochastic rate constant can be also related to the classical rate constant $k_i$ by $c_i=k_iV^{1-\nu}$, where, $\nu$ is the state change parameter or stoichiometric ratio of the reaction. Further, a particular jth reaction out of $M$ reaction channels will get fired if it satisfies the condition, $\displaystyle\sum_{i=1}^{j}\le a_0r_2<\sum_{i=1}^{j+1}$. Now, this SSA which incorporated these two fundamental concepts can able to simulate the trajectory of each molecular species in the system.\\

{\noindent}The degree of the fluctuations $\eta$ associated with dynamics of each molecular species trajectory can be estimated from the system size $V$ by the relation \cite{gillespie2000chemical}, $\eta\approx\frac{1}{\sqrt{V}}$. This relation indicates that the degree of the fluctuations inversely proportional to the square root of the system size $V$. In order to understand physical reality of this parameter, we can express $V$ as $V=N\times\frac{V}{N}=\frac{N}{\rho}$, where, $\rho=\frac{N}{V}$ is the density of the population in the demographic region. Since the population of the people in the demographic region is quite large as compared to the changes in the population in small time scale due to birth and death processes during interaction, one can take $N$ is approximately constant. This approximation gives us $\eta\sim\frac{1}{\sqrt{N}}\rho^{1/2}$, which provides us $\eta\propto\sqrt{\rho}$. In our study, we generated 30 independent stochastic realizations generated by the SSA and took the mean trajectory for model calibration and all nonlinear dynamical analyses presented in this study.\\

\begin{table*}[ht]
\centering
\caption{\label{Table1} Rate parameter values used in the dengue transmission model.}
\begin{tabular}{|c|c|c|c|c|c|}
\hline
\textbf{Parameter} & \textbf{Description} & \textbf{Used Value} ($\text{day}^{-1}$) & \textbf{Standard Value}  & \textbf{Reference} \\
\hline
$\lambda_h$, $\mu_h$ & Host per capita birth rate \& natural mortality rate&0.00003914 & $\dfrac{1}{70}\ \mathrm{year}^{-1}*$
& assumed \\
$\beta_{h0}$ & Vector to host transmission rate & 0.40 &$(0.03$--$0.75)~\mathrm{day}^{-1}$  &\cite{Andraud2012} \\
$\delta_h$ & Host progression rate  & 0.18333  & $(0.14$--$0.25)~\mathrm{day}^{-1}$ &\cite{ChanJohansson2012} \\
$\gamma_h$ & Host recovery rate &  0.075&$(0.07$--$0.14)~\mathrm{day}^{-1}$  &\cite{WHO2025DengueFactSheet} \\
$\lambda_v$, $\mu_v$ & Vector per capita birth rate \& natural mortality rate& 0.07143 & $(0.05$--$0.10)~\mathrm{day}^{-1}$   &\cite{Silva2020} \\
$\beta_{v0}$ & Host to vector transmission rate & 0.45 &  $(0.15$--$1.0)~\mathrm{day}^{-1}$  & \cite{Andraud2012}\\
$\delta_v$ & Vector progression rate  & 0.10 & $(0.083$--$0.125)~\mathrm{day}^{-1}$  & \cite{WHO2025DengueFactSheet}\\
\hline
\end{tabular}
\vspace{0.5em}
\begin{flushleft}
$*$ Average human host life span $\approx 70$ years.
\end{flushleft}
\end{table*} 

\begin{table}[h!]
\label{table1}
\centering
\caption{\label{Table2} Reaction networks and corresponding propensity functions for the stochastic SEIR--SEI dengue model}
\begin{tabular}{|c|c|}
\hline
\textbf{Reaction Network}     &        \textbf{Propensity Function } \\
\hline
$\emptyset \xrightarrow{\displaystyle B_h} S_h$ & $B_h$ \\
\hline
$S_h \xrightarrow{\displaystyle\beta_{h0} \frac{I_v}{N_h}} E_h$ & $\beta_{h0} \dfrac{S_h I_v}{N_h}$ \\
\hline
$E_h \xrightarrow{\displaystyle\delta_h} I_h$ & $\delta_h E_h$ \\
\hline
$I_h \xrightarrow{\displaystyle\gamma_h} R_h$ & $\gamma_h I_h$ \\
\hline
$S_h \xrightarrow{\displaystyle\mu_h} \emptyset$ & $\mu_h S_h$ \\
\hline
$E_h \xrightarrow{\displaystyle\mu_h} \emptyset$ & $\mu_h E_h$ \\
\hline
$I_h \xrightarrow{\displaystyle\mu_h} \emptyset$ & $\mu_h I_h$ \\
\hline
$R_h \xrightarrow{\displaystyle\mu_h} \emptyset$ & $\mu_h R_h$ \\
\hline
$\emptyset \xrightarrow{\displaystyle B_v} S_v$ & $B_v$ \\
\hline
$S_v \xrightarrow{\displaystyle\beta_{v0}\frac{I_h}{N_h}} E_v$ & $\beta_{v0} \dfrac{S_v I_h}{N_h}$ \\
\hline
$E_v \xrightarrow{\displaystyle\delta_v} I_v$ & $\delta_v E_v$ \\
\hline
$S_v \xrightarrow{\displaystyle\mu_v} \emptyset$ & $\mu_v S_v$ \\
\hline
$E_v \xrightarrow{\displaystyle\mu_v} \emptyset$ & $\mu_v E_v$ \\
\hline
$I_v \xrightarrow{\displaystyle\mu_v} \emptyset$ & $\mu_v I_v$ \\
\hline
\end{tabular}
\end{table}

\begin{table*}[ht]
\label{table2}
\centering
\caption{Calibrated transmission-modulation and observation parameters used in the stochastic SEIR--SEI model for each outbreak year.}
\label{tab:calibration}

\begin{tabular}{|l|l|c|c|c|}
\hline
\textbf{Parameter} & \textbf{Description} & \textbf{2023} & \textbf{2024} & \textbf{2025} \\
\hline
$\eta$ & Seasonal forcing amplitude & 0.10 & 0.10 & 0.30 \\
$T_{50}$ & Transmission activation time (days) & 210 & 225 & 190 \\
$n$ & Activation steepness & 6 & 6 & 6 \\
$\beta_{\rm floor}$ & Minimum transmission level & 0.032 & 0.032 & 0.035 \\
$\rho_{\rm imp}$ & Importation rate of exposed individuals & 0.026 & 0.022 & 0.025 \\
$\rho$ & Reporting factor & 1.00 & 1.05 & 1.30 \\
\hline
\end{tabular}
\end{table*}

\subsection{Extension of the model to calibrate with real data}
{\noindent}The validation of the stochastic $SEIR-SEI$ model simulation results with the collected real data as described in \ref{data} is important for further analysis of various dengue transmission dynamical properties and prediction. 
The preliminary simulation results of the stochastic reaction model of the baseline $SEIR-SEI$ given in Table \ref{table2} with the parameter values provided in Table \ref{table1} using SSA could not capture the dynamics of the real collected data of the dengue outbreaks. This could be due to the fact that the parameter values in the baseline model are taken as constants and do not explicitly account for temporal variation driven by various factors, such as, environmental conditions, seasonal vector abundance, low-level imported infections that influence dengue transmission dynamics. In view of this limitation, we have extended the model framework taking into account a few of the parameters involved in the model Table \ref{table1}. Since the transmission rates are key parameters in the dengue outbreak at any demographic region, we took two transmission rates involved in the stochastic $SEIR-SEI$ model, namely, human-to-vector ($\beta_h$) and vector-to-human ($\beta_v$) transmission rates to introduce time dependent periodic force \cite{kim2017assessment}. \\

{\noindent}The transmission rates in the general dengue transmission model are basically dependent on mainly two factors, seasonal modulating factor and transmission activation function \cite{polwiang2015seasonal,kim2017assessment}. In our proposed stochastic dengue transmission model. we considered these two driving factors in the transmission rates as follows. Let us define a periodic function $F(t)$ which represents a time dependent seasonal variability in the transmission rate parameters in dengue outbreak \cite{andraud2013simple},
\begin{equation}
\label{seas_forcing}
F(t)=1+\eta\cos\left(\frac{2\pi t}{365}\right),
\end{equation}
where $\eta$ denotes the amplitude of the seasonal periodic forcing which can be taken $0<\eta<1$ for dengue transmission dynamics \cite{andraud2013simple}. This function accounts for the seasonal modulation depending on the environmental conditions and climate that influence mosquito breeding, abundance and transmission intensity throughout the year \cite{andraud2013simple}. \\

{\noindent}Next, in order to model the simulated unbounded data into smooth distribution which may help in designing the increase in transmission activity and systematic analysis of the data in prediction and control of the dengue outbreak, we introduce a nonlinear activation sigmoid function $\sigma$ \cite{al2026modelling}. This function can be well approximated to the Hill response function \cite{gadagkar2015computational,skiadopoulos2025optimal} as given below,
\begin{equation}
\label{sigmoid}
\sigma(t)=\frac{t^n}{t^n+T_{50}^{n}},~~~t>0,
\end{equation} 
where, $T_{50}$ represents the characteristic activation time at which transmission reaches half of its maximum level and Hill coefficient $n$ controls the steepness of the transition from low to high transmission intensity rather than an abrupt increase \cite{goutelle2008hill}. Since the dengue transmission is rarely completely absent even during periods of low incidence, we also incorporated a minimum transmission level, which is a slightly modified functional form given in the reference \cite{andraud2013simple}, defined by the following function, 
\begin{equation}
\label{trans_lev}
A(t)=\beta_{\mathrm{floor}}+\left(1-\beta_{\mathrm{floor}}\right)\sigma(t),
\end{equation} 
where, \(\beta_{\mathrm{floor}}\) denotes the background transmission activity maintained throughout the outbreak period. Now, incorporating the equations \eqref{seas_forcing}, \eqref{trans_lev}, the time-dependent transmission rates for human-to-vector and vector-to-human transmission are given by the following equations,
\begin{eqnarray}
    \label{trans_rate}
    \left.
    \begin{aligned}
        \beta_h(t)=\beta_{h0}F(t)A(t),\\
        \beta_v(t)=\beta_{v0}F(t)A(t).
    \end{aligned}
    \right\rbrace
\end{eqnarray}
We incorporated these modified transmission rates into the stochastic $SEIR-SEI$ model and use them in all subsequent simulations. We did the simulation of the model using SSA, and the calibration of the simulation results were done independently on the collected outbreak data of the years, 2023, 2024 and 2025 respectively. The calibration parameters, including $\eta$, $T_{50}$, $n$, reporting factor $\rho$ and $\beta_{\mathrm{floor}}$ are adjusted iteratively to obtain the best agreement between the simulated and reported outbreak trajectories while maintaining biologically plausible epidemic behavior. Now, in order to compare with the reported incidence data, we generated 100 independent stochastic realizations of the model using SSA for each parameter set and took ensemble-average of the realizations. We then calculated the root mean square error ($R^2$-value) or relative error $(RE)$ between the simulated and dengue outbreak data for calibration and found that $R^2$-values are between $0.95<R^2<0.978$ which is in good agreement between the simulated and real outbreak data (Figure 2). We terminated the calibration procedure when further parameter adjustments produced negligible improvement in these performance measures. The final calibrated transmission-modulation parameters obtained for the 2023, 2024, and 2025 outbreaks are summarized in Table~\ref{tab:calibration} and these are subsequently used in the sensitivity analysis and all other nonlinear dynamical analyses presented in this study.

\subsection{Technique for sensitivity analysis}
{\noindent}The time series data we have, either dengue transmission real data or simulated data, are generally nonstationary data type which are very complex. The time series analysis techniques could able to dig patterns, sequential dynamics, segmentation etc which help future forecasting and control mechanisms \cite{brockwell2009time}. We briefly discuss few of the time series analytical techniques which we used to analyse our dengue transmission data. \\

{\noindent}One of the most important analysis of the dengue outbreak data (both real and model simulated data) is to identify the transmission-modulation parameters from the time series data that most strongly influence model performance. This analysis can be done using local sensitivity analysis (LSA) technique which can be defined as the time series sensitivity analysis method used to assess stability of the data to identify the influential system calibration parameters linking to the associated disease transmission and modulation \cite{qin2023comparative}. Because in the evaluation of the robustness of the calibrated model, the calculated trajectory sensitivity using LSA reveals the variation in the dynamical response in the data relating to change in only a certain parameters in the data \cite{ju2011load}. In our analysis, we considered only the calibration parameters associated with the time-dependent transmission rates, such as, the seasonal forcing amplitude $(\eta)$, the activation time $(T_{50})$, the Hill coefficient $(n)$, and the baseline transmission level $(\beta_{floor})$. We, then, perturbed each parameter independently by $\pm 10\%$ and $\pm 20\%$ relative to its calibrated value, while all other parameters are kept fixed. Further, the stochastic simulations of the model system for each perturbed parameter set are performed using the same procedure as described during the model calibration. Then, we compared the simulated incidence trajectories obtained from each perturbation with the reported dengue incidence data for each of the corresponding outbreak year. We used the changes in $R^2$ and $RE$ to evaluate the calibrated model's sensitivity to parameter variations. This local sensitivity analysis assesses parameter robustness in the neighborhood of the calibrated solution and complements the calibration procedure by identifying the transmission-modulation parameters which have the strongest impact on model performance \cite{qin2023comparative}.

\subsection{Statistical complexity analysis method}
 
{\noindent} Investigation of the dynamical features and organization of the time series of reported and simulated incidence of dengue can be carried out using a probabilistic based complexity analysis method at a given scale of a given dynamical system \cite{lopez1995statistical}. This technique characterizes the balance between randomness and order in the temporal structure of a time series to capture intrinsic patterns and associated information embedded in the time series and has been widely applied to the study of statistical properties of nonlinear dynamical systems \cite{feldman1998measures}. Brief description of the method is as follows \cite{rosso2007distinguishing}. Consider a time series defined by a time sequence $x_t$. Then, for a given $x_t$, ordinal patterns are constructed using an embedding dimension $d_x$ and time delay $\tau$. For each sequence of $d_x$, consecutive inequality observations are transformed into a permutation of these inequalities according to the relative ordering of its value \cite{bandt2002permutation}. The probability distribution of all admissible ordinal patterns is then estimated from the time series. This allows us to calculate the normalized permutation entropy given by \cite{bandt2002permutation},
\begin{equation}
\label{perm_entr}
H=-\frac{1}{\log_2(d_x!)}\sum_{i=1}^{d_x!} p_i \log_2 p_i,
\end{equation}
where, $p_i$ denotes the probability associated with $i^{th}$ ordinal pattern. The quantity $H$ ranges between 0 and 1, with larger values indicating greater dynamical randomness \cite{bandt2002permutation,riedl2013practical}. Then, this $H$ can be used to measure statistical complexity or entropic non-triviality measure $C$ of the time series data based on the extensive Jensen-Shannon divergence technique \cite{lopez1995statistical,rosso2007distinguishing},
\begin{equation}
C=Q_J(P,P_e)H,
\end{equation}
where, $Q_J(P,P_e)$ is the disequilibrium parameter which can be calculated from the the normalized Jensen-Shannon divergence between the observed ordinal pattern distribution $P$ and the uniform distribution $P_e$ \cite{rosso2007distinguishing,lamberti2004intensive}. This statistical complexity $C$ quantifies both randomness and the degree of correlational structural organization present in the time series \cite{lopez1995statistical,lamberti2004intensive}. In this technique, the statistical complexity or complexity-entropy $(CH)$ measure causality plane is defined as the two-dimensional plane of permutation entropy $H$ and statistical complexity measure $C$ which is known as $CH-plane$. Provided $0\le H\le 1$, $C$ is bounded with $C_{min}\le C\le C_{max}$ and these values characterize the properties of the dynamics, such as, chaos, stochastic noise etc \cite{rosso2007distinguishing}. In our study, the positions occupied by the reported and simulated incidence series in the $CH$-plane are used to compare their dynamical characteristic and determine how well the calibrated stochastic model reproduces the organization of the observed outbreak dynamics.\\

{\noindent}Further, to examine temporal variations in the system's dynamical behavior, we used a sliding-window analysis of permutation entropy, where, each window's entropy value is calculated independently, generating a time-dependent entropy series that characterizes changes in the dynamical organization of the reported and simulated incidence records throughout the study period \cite{bandt2002permutation}. This analysis enables us to track the changes in the complexity of the epidemic dynamics throughout each outbreak. Further, for each embedding delay, the Euclidean distance between the reported and simulated model stochastic time series in the $CH-$plane is calculated for each valid sliding window as given by \cite{da2024symbolic},
\begin{equation}
\label{ch_distance}
D_k =
\sqrt{
\left(H_k^{\mathrm{rep}}-H_k^{\mathrm{SSA}}\right)^2+
\left(C_k^{\mathrm{rep}}-C_k^{\mathrm{SSA}}\right)^2
},
\end{equation}
where, \(H_k\) and \(C_k\) denote the permutation entropy and statistical complexity, respectively, and \(k\) denotes the sliding-window index. The mean complexity--entropy distance is then calculated as,
\begin{equation}
\label{mean_ch_distance}
\overline{D}_{CH}
=
\frac{1}{N}
\sum_{k=1}^{N} D_k,
\end{equation}
where, \(N\) is the number of valid sliding windows. The embedding delay yielding the minimum \(\overline{D}_{CH}\) is selected independently for each outbreak year and is subsequently used for the final complexity--entropy analysis.

\subsection{Recurrence quantification analysis method}

{\noindent}The temporal organization of patterns and their repetitive phenomena are some of the important characteristics of the natural time series \cite{marwan2007recurrence}. These properties can be captured and studied using \textit{recurrence plot or recurrence quantification analysis (RQA)} and this method is important specially for time series analysis \cite{eckmann1995recurrence}. This $RQA$ is a method which can be used to measure embedded repetitive patterns and hidden dynamical changes by quantifying the recurrence properties, persistence, and predictability of the underlying dynamical system in the reconstructed phase space \cite{marwan2007recurrence,webber2015recurrence}. Within this framework, for a given reconstructed phase-space trajectory $\mathbf{x}_i$, RQA is defined by the recurrence matrix given by \cite{marwan2007recurrence},
\begin{equation}
\label{rec_mat}
R_{ij}=\Theta\left(\varepsilon-\left\|\mathbf{x}_i-\mathbf{x}_j\right\|\right),
\end{equation}
where, $\Theta(\cdot)$ denotes the Heaviside function, $\varepsilon$ is the recurrence threshold, and $\|\cdot\|$ represents the Euclidean distance. This recurrence matrix identifies pairs of states that are sufficiently close in the reconstructed phase space and it is visualized through recurrence plots. Hence, RQA is a graphical representation of the repeated visits to similar dynamical states and associated dynamical changes in the time series data \cite{webber2015recurrence}. The structured diagonal and vertical line patterns in the recurrence plots indicate persistent and deterministic dynamical behavior, respectively \cite{marwan2007recurrence}.\\

{\noindent}Further, the RQA method can be used to quantify important recurrence structures of a complex time series data, such as, determinism $(DET)$, laminarity $(LAM)$, and trapping time $(TT)$ \cite{webber1994dynamical,marwan2002recurrence}. Determinism measures the proportion of recurrence points forming diagonal line structures and is associated with predictable temporal evolution \cite{webber1994dynamical}. Laminarity quantifies the occurrence of vertical line structures and reflects intermittent or slow varying behavior \cite{marwan2002recurrence}. Trapping time represents the average length of vertical structures and characterizes the persistence of laminar states \cite{marwan2002recurrence}. We compute recurrence plots and $RQA$ measures for both collected and simulated time series data of dengue outbreak. We also did comparative analysis of these quantities to assess the extent to which the calibrated stochastic model reproduces the recurrence organization observed in the real collected outbreak dynamical data.

\section{results}
\label{sec:results}
{\noindent}We present the results of our investigation on dengue outbreak in Imphal, Manipur for the three consecutive years 2023, 2024 and 2025 respectively. We used the collected data of dengue transmission cases of the region and years we mentioned as discussed in the method \ref{data}. The simulations of the proposed stochastic SEIR-SEI model are done using SSA using the reaction channels given in the Table \ref{Table1} with the parameter values given in Table \ref{Table2}.

\subsection{Stochastic model simulation results well calibrated with the Dengue outbreak data}
\begin{figure*}
    \centering
    \includegraphics[scale=0.5]{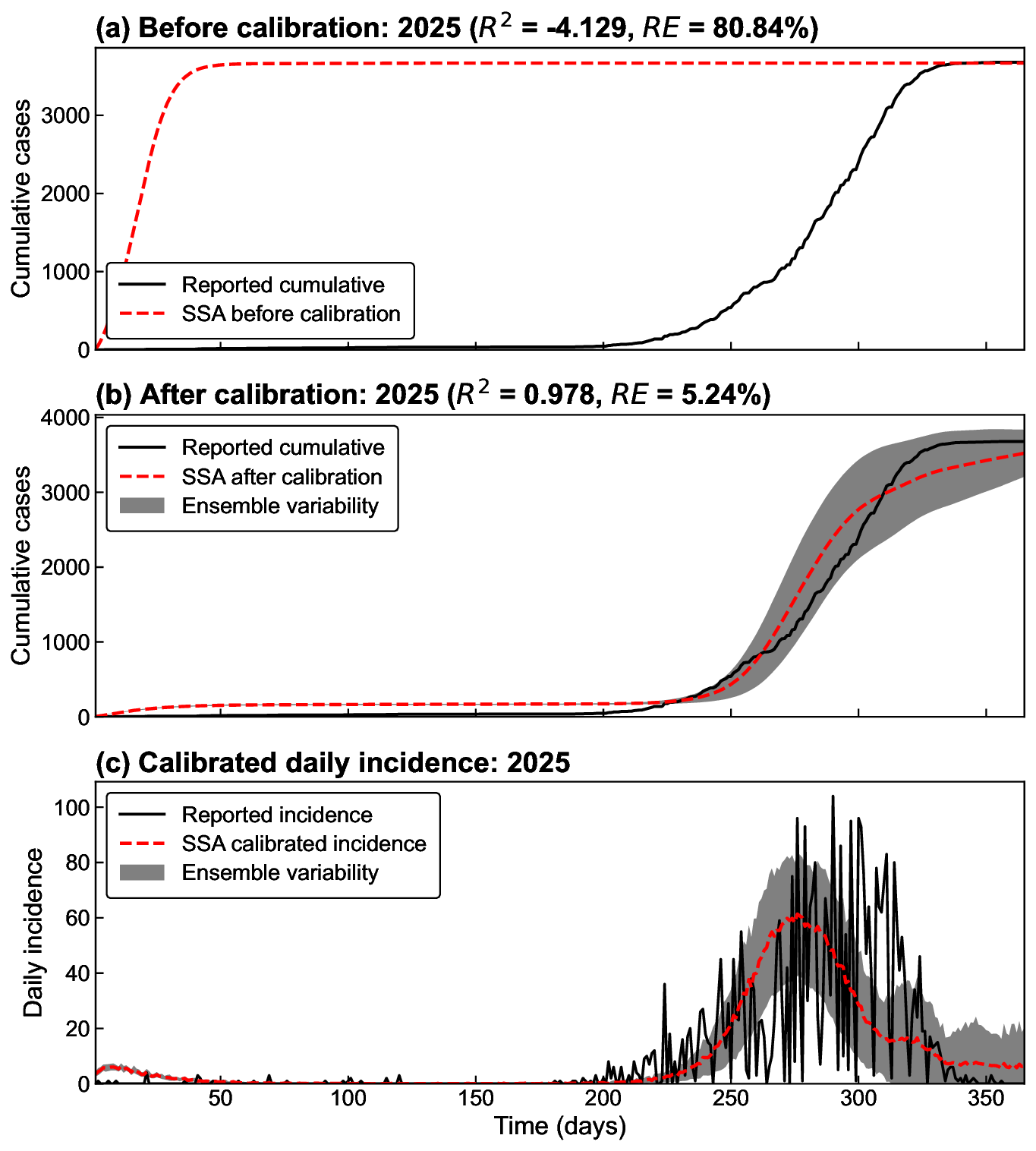}
    \caption{ Calibration of the stochastic SEIR--SEI model for the 2025 dengue outbreak. \textbf{(a)} Comparison between the reported cumulative incidence and the stochastic simulation prior to calibration. \textbf{(b)} Reported cumulative incidence together with the calibrated stochastic simulation. \textbf{(c)} Comparison between the reported and simulated daily incidence trajectories obtained from the calibrated model. The shaded region represents the variability obtained from 30 independent stochastic realizations, illustrating the uncertainty associated with the stochastic transmission process.}
    \label{fig2}
\end{figure*}

\begin{figure*}
    \centering
    \includegraphics[scale=0.6]{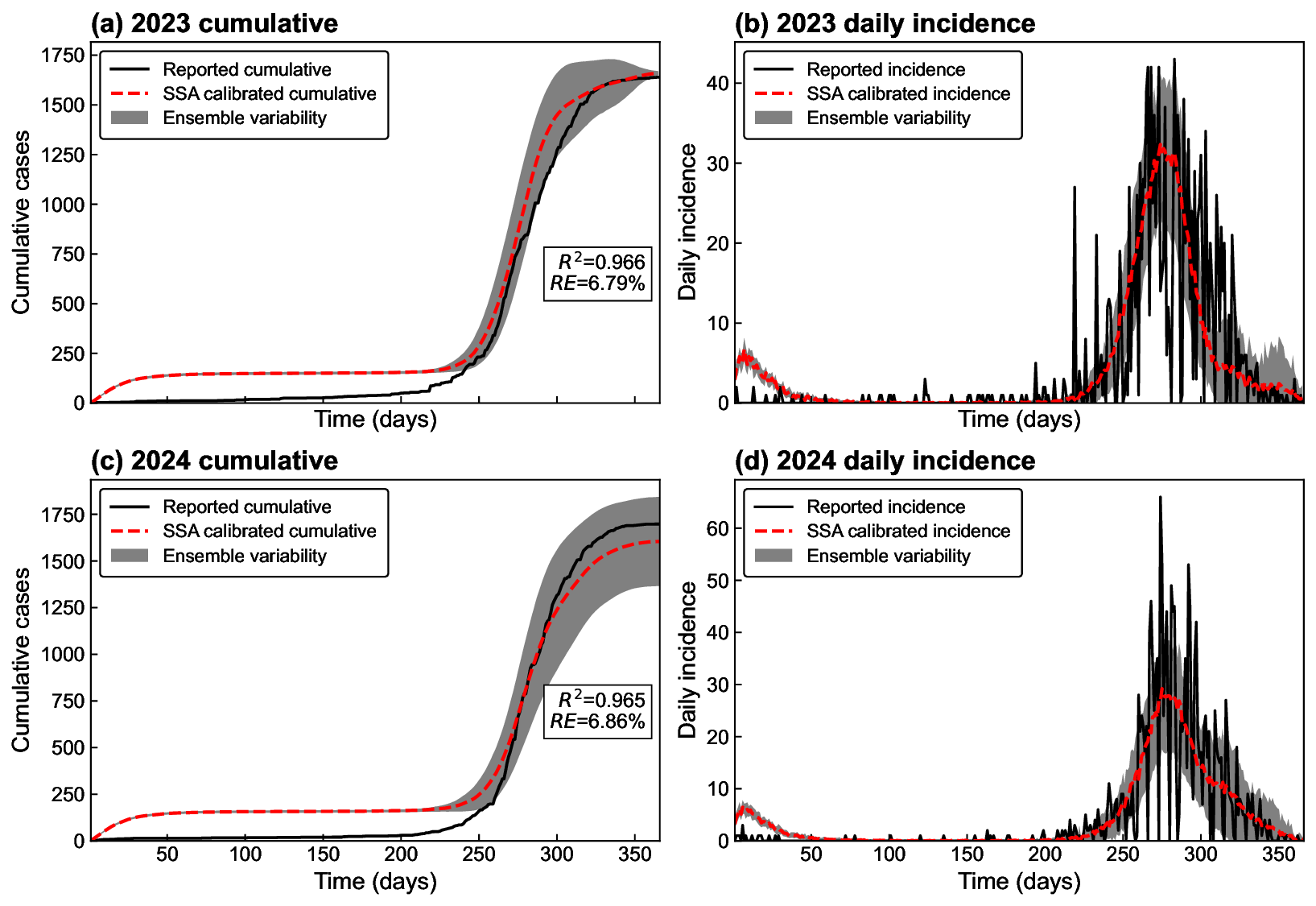}
    \caption{ Independent calibration of the stochastic SEIR--SEI model for the 2023 and 2024 dengue outbreaks. The calibrated simulations reproduce the reported cumulative [\textbf{(a)} \& \textbf{(c)}] and daily incidence trajectories [\textbf{(b)} \& \textbf{(d)}] for both outbreak years, capturing the onset, epidemic growth, peak timing, and decline of each outbreak. The shaded regions denote the variability obtained from 30 independent stochastic realizations.}
    \label{fig3}
\end{figure*}

{\noindent}We simulated our stochastic SEIR-SEI model defined by the set of reactions given in the Table \ref{Table1} with the parameter values provided in the Table \ref{Table2} as well as constant and modified time dependent transmission rates given by the equations \eqref{seas_forcing}, \eqref{sigmoid}, \eqref{trans_lev} and \eqref{trans_rate} using SSA. From the results, one can observe that the SSA results with constant transmission rates (represented by dash red line) could not able to properly capture dynamics of the observed cumulative incidence (represented by the black solid curve) dengue outbreak data (Fig. \ref{fig1}(a)) as evident by the values of $R^2=-4.129$ and $RE=80.84\%$. However, if the modified time dependent transmission rates are considered, the simulation results are now calibrated with the collected dengue transmission data of 2025 outbreak as shown in Fig. \ref{fig2} (b) and (c), and the model simulation results could able to capture significantly well the observed seasonally driven temporal evolution of the outbreak ($R^2=0.978$ and $RE=5.24\%$). Further, the model results also accurately validate the outbreak onset, the epidemic growth phase, the timing of the incidence peak, and the subsequent decline while preserving the stochastic variability (represented by the grey shaded region) of the transmission process. Similarly, we also present the calibrated results for the 2023 and 2024 outbreaks in Fig.\ \ref{fig3}. In both years, the ensemble-average stochastic simulations are closely matched with the reported cumulative and incidence dynamics throughout the outbreak period (Fig.\ \ref{fig3}(a) \& (c)). Although there were minor differences in the magnitude of individual daily fluctuations (Fig.\ \ref{fig3}(b) \& (d)), the proposed model could able to reproduce the main epidemiological characteristics of collected dengue outbreak data of each year, including the timing of epidemic onset, growth, peak transmission, and decline. This agreement between the model simulation data and real collected dengue transmission data is evident from the coefficients of determination of $R^2=0.966$ and $0.965$ for the 2023 and 2024 outbreaks, respectively, with corresponding $RE=6.79\%$ and $RE=6.86\%$. These consistently high goodness-of-fit values demonstrate that the calibrated stochastic model provides a robust representation of the reported dengue outbreaks across three consecutive transmission seasons.\\

\begin{figure*}
    \centering
    \includegraphics[scale=0.6]{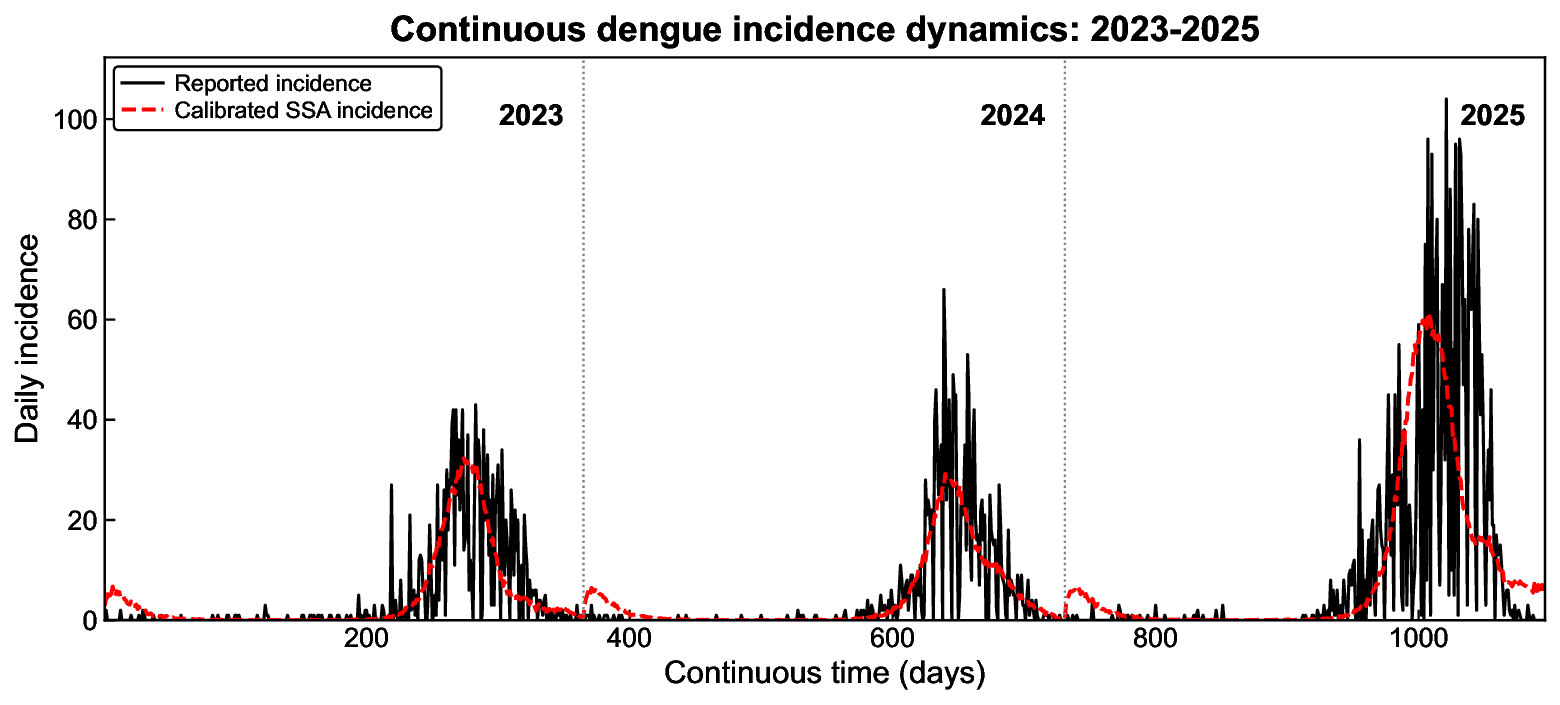}
    \caption{Continuous reported and calibrated simulated dengue incidence series constructed by concatenating the independently calibrated outbreaks from 2023, 2024, and 2025. The resulting three-year time series preserves the temporal characteristics of each annual outbreak.}
    \label{fig4}
\end{figure*}

{\noindent}Next, in order to confirm that our proposed stochastic model can capture the quantitative behavior of dengue outbreak, we calibrated our stochastic model simulation data with the daily dengue incidence record data for the continuous three outbreak years 2023-2025 (Fig.\ \ref{fig4}). This model validation with the recorded continuous dataset is found to be closely in agreement. The close agreement between the reported and simulated incidence data provides a reliable foundation for investigating whether the calibrated model also reproduces the higher-order dynamical organization of the observed dengue outbreaks. This means that the proposed model can retain the temporal characteristics of individual outbreaks as well as longer incidence record of the dengue outbreak, and hence, the proposed model is suitable for the analysis of the dynamical properties of the outbreak, such as, transmission patterns, prediction of epidemic/pandemic epicenters and endemic, control of the transmission etc. 


\subsection{Sensitivity analysis of calibration parameters}
\begin{figure*}
    \centering
    \includegraphics[scale=0.7]{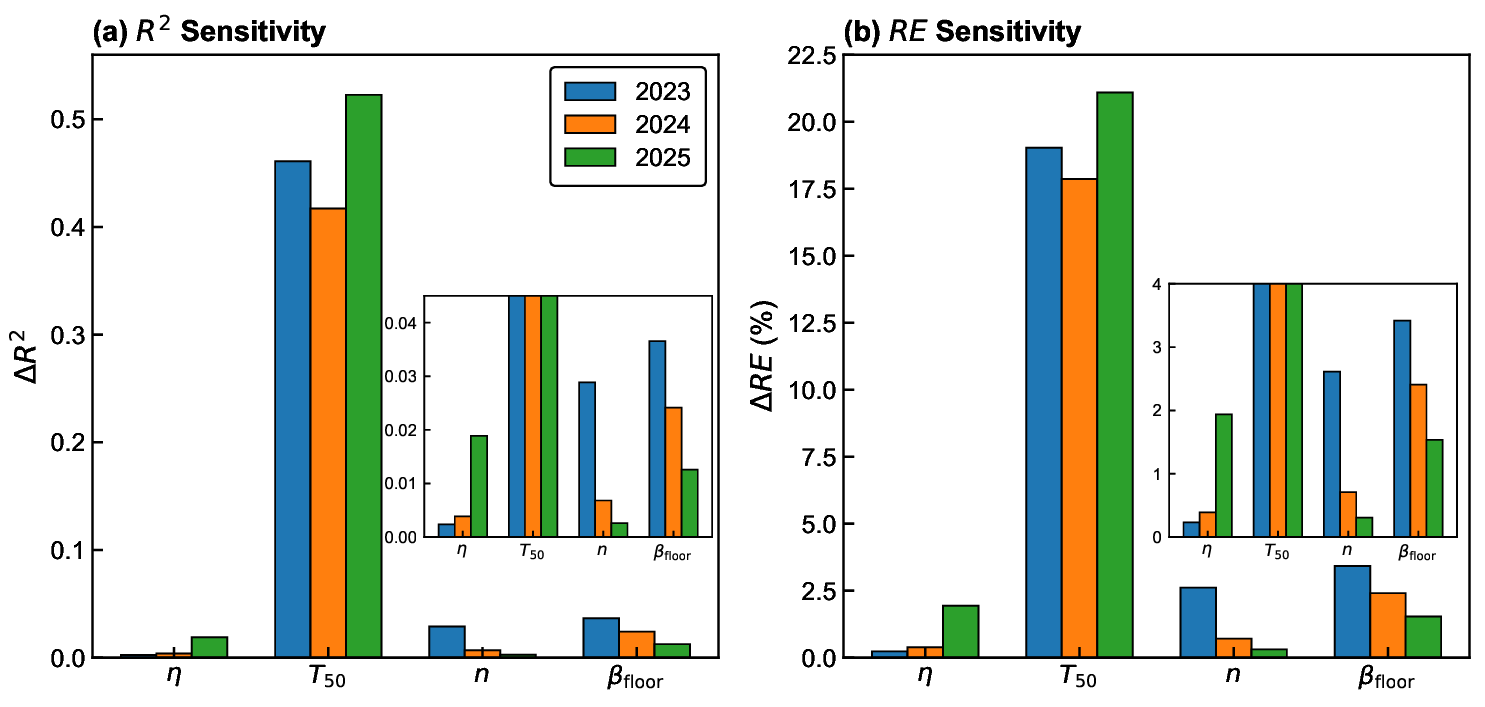}
        \caption{ Sensitivity analysis of the transmission-modulation parameters. Relative changes in the \textbf{(a)} coefficient of determination $R^2$ and \textbf{(b)} relative error (RE) resulting from parameter perturbations around the calibrated values. The results demonstrate that the calibration performance is most sensitive to variations in the activation time $T_{50}$, while the seasonal forcing amplitude $(\eta)$, activation steepness (n), and baseline transmission level $\beta_{\mathrm{floor}}$ exert comparatively smaller effects.} 

    \label{fig5}
\end{figure*}

{\noindent}We examine the robustness of the calibrated stochastic SEIR--SEI model by evaluating the response of the calibration metrics to independent perturbations of the transmission-modulation parameter (Fig.\ \ref{fig5}). The results show that the proposed model is performed consistently well under the moderate model parameter variations, indicating that the calibrated solution is locally robust within the neighborhood of the optimal parameter set. The transmission activation time $T_{50}$ showed the greatest influence on model performance. The observed largest variation in $R^2$, and the relative error for all the three outbreak years from their corresponding calibrated values reveal the strongest response observed for the 2024 outbreak. However, the perturbations in the other model parameters, such as, the seasonal forcing amplitude $(\eta)$, the hill coefficient $(n)$, and the minimum transmission level $\beta_{\rm floor}$ resulted in minimal changes in both calibration metrics. This observed results indicates that the model performance is relatively insensitive to moderate variation in these parameters.\\

{\noindent}On the other hand, the predominance of $T_{50}$ emphasizes the importance of accurately capturing the temporal onset of transmission during outbreak development. Moreover, even moderate deviation in $T_{50}$, which determines the transition from the low-transmission phase to the period of sustained epidemic growth, can alter the timing of epidemic expansion, leading to a measurable deterioration in agreement with the reported incidence data. In comparison, moderate changes in the values of the remaining transmission-modulation parameters have a larger impact on transmission magnitude than the temporal organization, resulting in only minor changes in overall calibration performance. Taken together, these results show that the calibrated model is locally robust, with the timing of transmission activation being the primary determinant of calibration accuracy.
 
\subsection{Complexity--Entropy Structure of the Outbreak Dynamics}

\begin{figure*}
    \centering
    \includegraphics[scale=0.45]{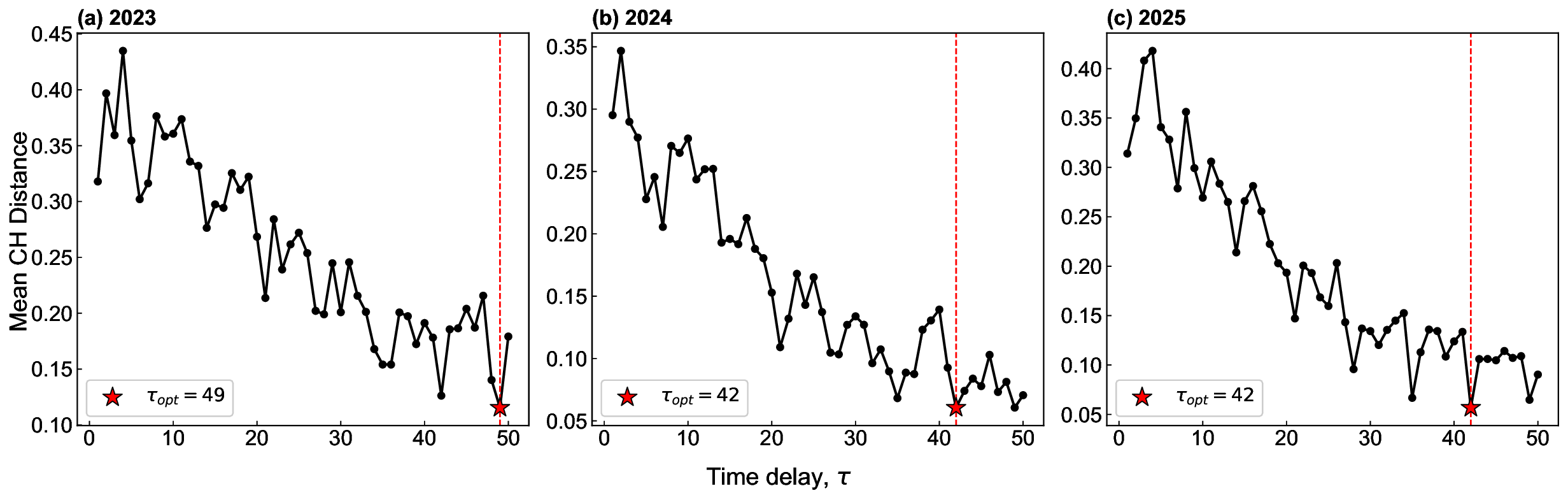}
    \caption{Year-wise optimization of the embedding delay used in the complexity--entropy analysis.
    The mean distance between the reported and calibrated stochastic series in the CH plane is evaluated over the tested delay range.
    The optimal delays are $\tau=49$, $42$, and $42$ for 2023, 2024, and 2025, respectively.}
    \label{fig6}
\end{figure*}

\begin{figure*}
    \centering
    \includegraphics[scale=0.45]{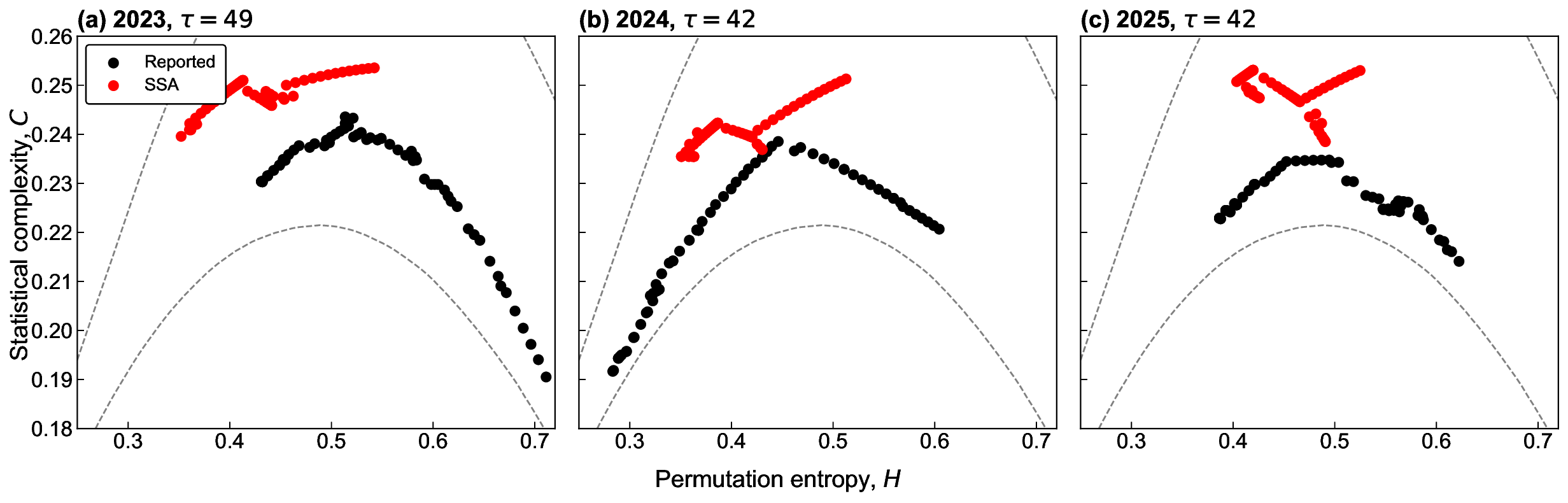}
    \caption{ Year-wise complexity--entropy planes of the reported and calibrated stochastic dengue incidence series evaluated at the corresponding optimal embedding delay.
    Panels \textbf{(a)}, \textbf{(b)}, and \textbf{(c)} correspond to 2023 ($\tau=49$), 2024 ($\tau=42$), and 2025 ($\tau=42$), respectively.
    Each point represents a sliding-window estimate of normalized permutation entropy $H$ and statistical complexity $C$.
    The dashed curves indicate the numerically estimated admissible CH boundaries.}
    \label{fig7}
\end{figure*}

{\noindent}We, next, study a year wise statistical complexity analysis using a sliding-window of the incidence time series data to examine whether the calibrated stochastic model reproduces the nonlinear dynamical organization of the reported dengue outbreaks beyond the agreement achieved through conventional calibration metrics. This analysis is done by optimizing the embedding delay $\tau$ independently for each outbreak year by evaluating the mean distance between the reported and calibrated stochastic model simulation trajectories in the complexity-entropy plane (CH-plane) over the tested delay range of $1\leq\tau\leq50$ days. The resulting delay optimization curves are shown in Fig.~\ref{fig6}. The minimum mean CH distance occurs at $\tau=49$ for 2023 and at $\tau=42$ for both 2024 and 2025 respectively. Thus, the characteristic delay that provides the closest CH-space correspondence is not identical across the three outbreak years, indicating year to year differences in the temporal organization of the incidence records of the dengue outbreak. Using the year specific optimal embedding dalays, the corresponding statistical behavior of the outbreak are shown in the CH-plane which are given in Fig.~\ref{fig7}. The CH analysis is performed using a sliding window of 300 days and embedding dimension of $d_x=3$, with the embedding delay optimized independently over $1\leq\tau\leq50$ days for each year. The reported and calibrated stochastic series occupy partially overlapping regions of the admissible complexity-entropy domain, while retaining visible differences in their distribution within the plane.\\

{\noindent}Now, the detail results of the year wise characteristics of the dengue transmission complexity are discussed as follows. For 2023, the reported series exhibits a mean permutation entropy of $\langle H\rangle=0.543$ and a mean statistical complexity of $\langle C\rangle=0.230$. The corresponding calibrated stochastic series gives  $\langle H\rangle=0.430$ and $\langle C\rangle=0.249$, with the two series separated by a CH-space distance of $0.115$. This indicates that the calibrated trajectory occupies a lower-entropy and slightly higher-complexity region than the reported series.
For 2024, the reported and calibrated series show considerably closer agreement in permutation entropy, with $\langle H\rangle=0.416$ and $0.408$, respectively. Their corresponding mean statistical complexities are $\langle C\rangle=0.219$ and $0.241$, giving a CH-space separation of $0.023$. Thus, among the three years, the calibrated and reported series are most closely aligned in the CH plane during 2024.
Finally, for 2025, the reported series has $\langle H\rangle=0.490$ and $\langle C\rangle=0.226$, whereas the calibrated series gives $\langle H\rangle=0.452$ and $\langle C\rangle=0.248$. The resulting Ch-space separation is $0.044$, indicating an intermediate level of agreement level of agreement relative to 2023 and 2024. These results show that the CH-separation is year dependent, with the largest difference occurring in 2023 and the smallest in 2024. Across all three years, the calibrated series consistently occupies a somewhat lower entropy and higher-complexity region than the reported series, although the magnitude of this displacement varies substantially between years. The CH analysis, therefore, indicates that the calibrated stochastic model captures some of the observed temporal organization while retaining systematic differences in the ordinal pattern structure of the reported incidence.\\ 

\begin{figure*}
    \centering
    \includegraphics[scale=0.45]{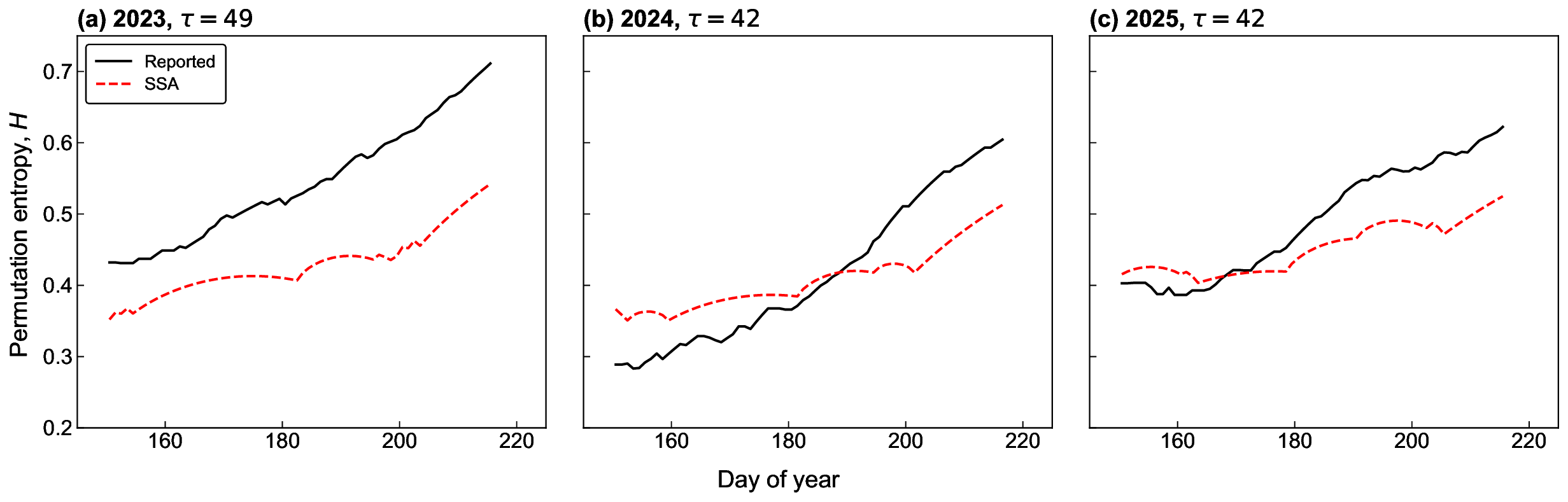}
    \caption{Temporal evolution of normalized permutation entropy $H$ for the reported and calibrated stochastic dengue incidence series.
    Panels \textbf{(a)}, \textbf{(b)}, and \textbf{(c)} correspond to 2023, 2024, and 2025, respectively, and use the year-specific optimal delays obtained from the CH-distance optimization.
    The horizontal axis represents the day of year corresponding to the center of each sliding analysis window.}
    \label{fig8}
\end{figure*}

{\noindent}We also give additional insights through the sliding-window evolution of permutation entropy in Fig.\ \ref{fig8}. The entropy curve of both the reported and calibrated series vary systematically across the analysis windows, demonstrating that the dynamical organization changes during the course of each outbreak. The calibrated time series follows several of the broad temporal trends present in the reported incidence, particularly, during the transition toward the later stages of the outbreak, but its entropy trajectory does not coincide with the empirical series. The differences are most apparent in 2023, where the reported increases to substantially higher values than the calibrated series toward the end of the analyzed interval. In 2024 and 2025, the two entropy trajectories remain closer, although systematic differences are still present.\\

{\noindent}Finally, the CH analysis shows that the good agreement between model simulation and recorded incidence data at the trajectory level does not necessarily imply identical ordinal-pattern organization. The calibrated stochastic time series occupies a broadly similar region in the CH-plane as the reported incidence data, while retaining systematic differences in entropy and statistical complexity across all the three years. The relatively small CH-separation in 2024 and 2025 indicate closer correspondence in their nonlinear temporal organization, whereas, the larger separation in 2023 points to a greater mismatch. These results demonstrate that the calibrated model captures substantial aspects of the observed temporal organization, but does not fully reproduce its higher order structure. The CH analysis provides a complementary diagnostic to conventional goodness-of-fit measures and motivates the subsequent recurrence based analysis of temporal organization.

\subsection{Recurrence structure of outbreak dynamics}
\begin{figure*}
    \centering
    \includegraphics[scale=0.7]{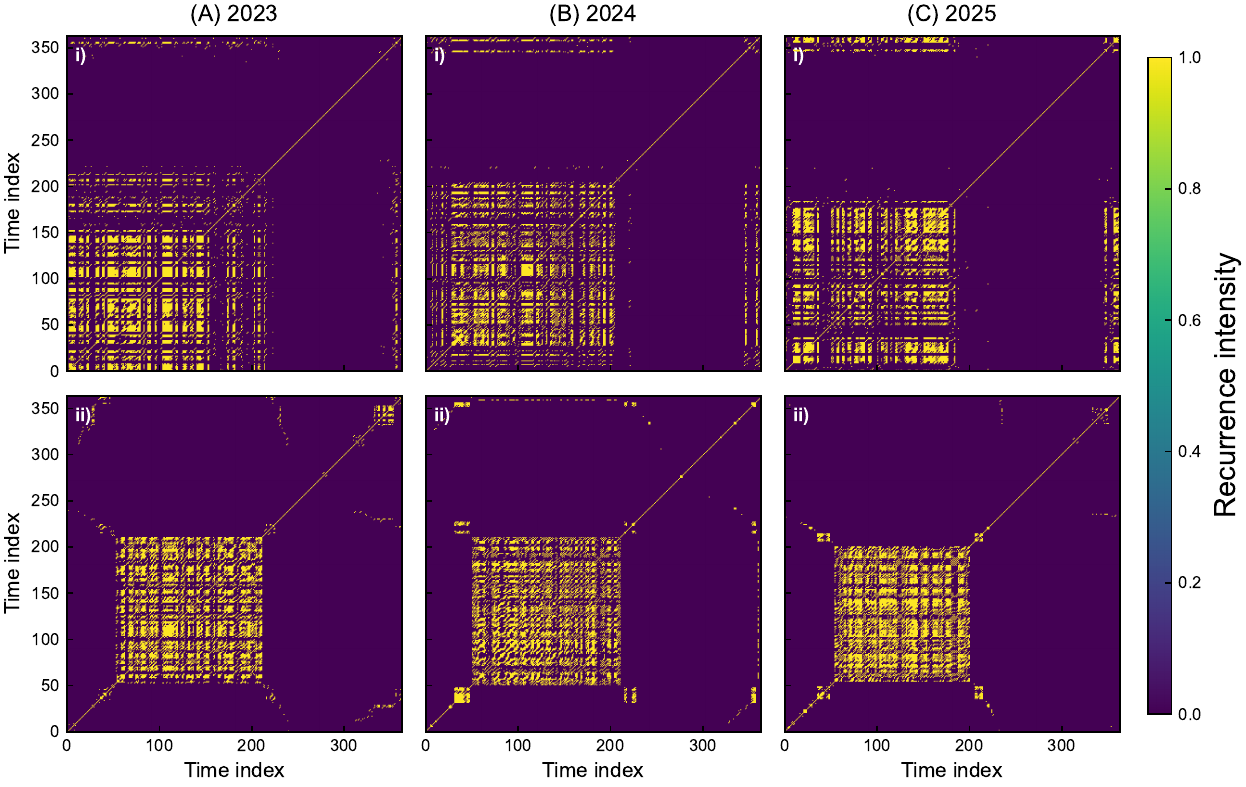}
    \caption{Year-wise recurrence plots of the reported and calibrated stochastic dengue incidence series reconstructed using embedding dimension $m=3$ and embedding delay $\tau=1$ day. Panels \textbf{(A)}, \textbf{(B)}, and \textbf{(C)} correspond to the 2023, 2024, and 2025 outbreaks, respectively. For each year, panel \textbf{(i)} shows the reported incidence and panel \textbf{(ii)} shows the calibrated stochastic simulation. The same recurrence-intensity scale is used throughout the figure.}
    \label{fig9a}
\end{figure*}

\begin{figure*}
    \centering
    \includegraphics[scale=0.7]{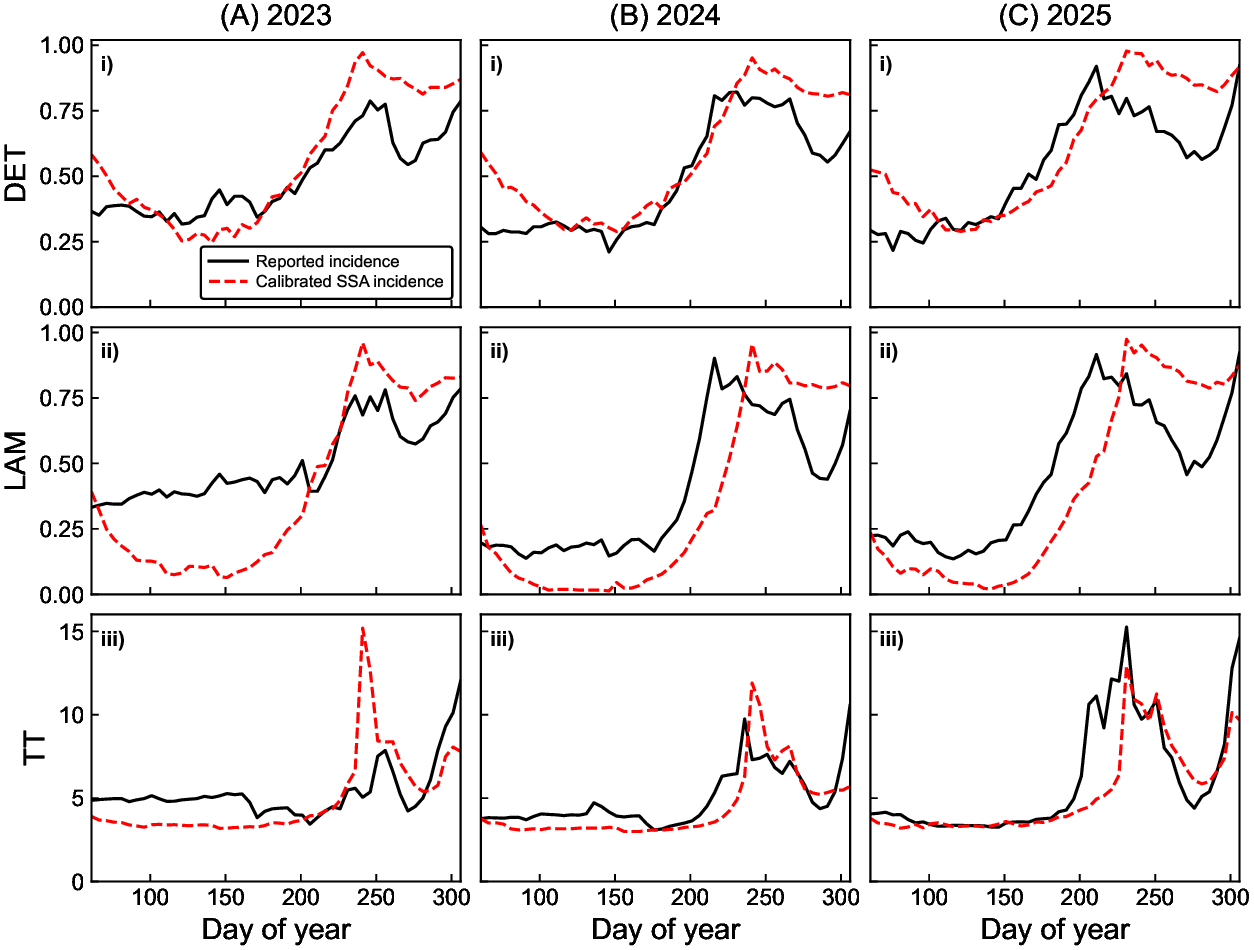}
    \caption{Year-wise temporal evolution of recurrence quantification measures for the reported and calibrated stochastic dengue incidence series. Panels \textbf{(i)}, \textbf{(ii)}, and \textbf{(iii)} show determinism (DET), laminarity (LAM), and trapping time (TT), respectively. The horizontal axis represents the day of year corresponding to the center of each sliding analysis window. The black solid and red dashed curves represent the reported and calibrated stochastic incidence series, respectively.}
    \label{fig9b}
\end{figure*}

{\noindent}We now apply recurrence quantification analysis (RQA) to further examine whether the calibrated stochastic model reproduces the temporal organization of the reported dengue outbreaks. We present the recurrence plots together with the temporal evolution of determinism (DET), laminarity (LAM), and trapping time (TT) as shown in Fig.\ \ref{fig9a} and Fig.\ \ref {fig9b} respectively. In this analysis, we used an embedding  dimension of \(m=3\) and a fixed embedding delay of \(\tau=1\) day for both the reported and calibrated stochastic time series data. A 120-day sliding window was advanced in 5-day steps, with a recurrence rate of 0.08 and minimum diagonal and vertical line lengths of \(l_{\min}=3\) and \(v_{\min}=3\) respectively. The recurrence plots reveal clear structured pattern in both the reported and calibrated stochastic time series data rather than a random distribution of recurrence points. The two representations, however, do not exhibit identical recurrence patterns. The calibrated simulations show more pronounced diagonal structures, while the reported incidence displays comparatively stronger vertical structures. These differences are also reflected in the temporal behavior of the RQA measures. The year wise analysis of the stochastic model simulation data and reported dengue outbreak data are presented below.\\

{\noindent}The analysis of 2023 data shows that the reported incidence has a mean determinism of \(0,497\), while the calibrated stochastic model simulation data also shows greater laminarity, with a mean value of \(0.499\), and a longer trapping time of \(5.341\) as shown in Fig.\ \ref{fig9a} (A) and Fig.\ \ref {fig9b} (B) respectively. The corresponding characteristics of the calibrated time series data are \(0.407\) for laminarity and \(4.909\) for trapping time. Thus, the calibrated model simulation data exhibits stronger diagonal recurrence organization, whereas, the reported incidence retains more persistent vertical recurrence structures.\\

{\noindent}A similar pattern as obtained in 2023 is observed in 2024. The mean determinism is \(0.486\) for the reported incidence data and increase to \(0.568\) in the calibrated model simulation data as shown in Fig.\ \ref{fig9a} (B) and Fig.\ \ref {fig9b} (B) respectively. The reported time series maintains higher laminarity, with a mean value of \(0.398\), and a trapping time of  \(4.921\).\\

{\noindent}The results of the 2025 data analysis indicated that the mean determinism reaches \(0.536\) for the reported time series data and \(0.609\) for the calibrated model simulation data as shown in Fig.\ \ref{fig9a} (C) and Fig.\ \ref {fig9b} (C) respectively. The reported incidence data exhibit stronger laminar recurrence, with a mean laminarity \(0.460\) and a trapping time of \(6.097\) respectively. The calibrated model simulation time series data gives laminarity value \(0.404\) and trapping time of \(5.339\) respectively. Notably, the trapping time of the reported incidence data is higher than the preceding years, indicating increased persistence of vertically organized recurrent states during this outbreak. \\

{\noindent}The analysis of the RQA of the reported and model simulation data of the three years gives us the following overall results. Across the three years, we found that the  the calibrated stochastic simulations data shows consistently exhibit larger determinism as compared to the reported incidence dengue outbreak data, whereas, this reported incidence data shows larger laminarity and longer trapping times as compared to the model simulation data. The temporal RQA profiles further indicate that these differences are not uniform throughout each outbreak. Both the representations undergo changes in recurrence organization with time, but the timing and magnitude of these changes also differ between the reported and calibrated series. However, the good agreement of the RQA results between the model simulation data and recorded incidence data does not correspond to the identical recurrence organization. The calibrated model simulation data captures substantial recurrent structure, but its recurrence dynamics differ systematically from those of the reported outbreaks data, particularly in the balance between diagonal and vertical recurrence structures. Combined with the CH analysis indicate that the trajectory-level agreement alone is insufficient to establish the equivalence in higher-order temporal organization.\\

{\noindent}In summary, the present results demonstrate that the proposed seasonally driven stochastic SEIR-SEI dengue transmission model, which incorporated the nonlinear dynamical effects, provides a good agreement with the recorded dengue incidence data as evident from the significantly large values of conventional goodness-of-fit measures, and carries quite similar dynamical properties and structures of embedded information. By combining with complexity entropy and recurrence quantification analyses, the present study validates not only the agreement between simulated and reported outbreak trajectories, but also their temporal complexity and recurrence organization. The results show that the calibrated stochastic model captures substantial features of the observed nonlinear temporal structural features, while systematic differences remain in the calculation of entropy, statistical complexity, and recurrence organization. Thus, the fluctuations, seasonally driven model parameters and nonlinear diagnostics provide an additional level of assessment beyond trajectory-level temporal characteristics and offer a more informative basis for evaluating the dynamical consistency with the proposed stochastic dengue outbreak model.

\section{Discussion}
\label{sec:discussion}
{\noindent}The general framework of constructing stochastic dengue outbreak model is strictly based on the proper validation of the model generated data with the real recorded incidence data available, where, significantly large number of realizations produced by the model simulation dat should be stably quite in agreement with the real data quantitatively. Although such calibrations provide an important measure of model performance, it offers only limited information about whether the simulated outbreak data replicate the underlying nonlinear dynamical organization of the observed dynamical outbreak characteristics. In the current study, this issue was addressed by combining conventional model calibration with complexity-entropy and recurrence analyses to evaluate the dynamical consistency of a calibrated stochastic SEIR--SEI model. This integrated framework extends model validation beyond goodness-of-fit by examining both epidemiological agreement and the temporal organization of dengue outbreak dynamics.\\

{\noindent}The incorporation of seasonally modulated transmission parameters together with a gradual transmission activation mechanism, significantly improved the agreement between simulated and reported outbreaks over the three consecutive years. The sensitivity analysis showed that the calibration solution is locally robust, with the transmission activation time $T_{50}$ having the largest impact on model performance. This finding emphasizes the importance of representing time-dependent transmission processes in the stochastic dengue outbreak model, as the timing of seasonal transmission activation has a significant impact on the temporal evolution of dengue outbreaks.\\

{\noindent}The complexity-entropy analysis provided an assessment of the dynamical organization of the outbreak time series in addition to conventional goodness-of-fit measures of the data. The reported and simulated incidence time series carry broadly a similar region of the complexity-entropy plane, indicating a degree of correspondence of agreement in their temporal organization. However, systematic differences remained in both permutation entropy and statistical complexity analysis. Across all three years, the calibrated stochastic model simulation data exhibited somewhat lower entropy and higher statistical complexity than the reported incidence data, although the magnitude of their difference varied in different years. The CH-space separation is largest in 2023 and substantially smaller in 2024 and 2025 respectively. These results indicate that the close agreement in the incidence trajectories does not necessarily imply identical ordinal-pattern organization. Rather, the CH analysis reveals residual differences in the temporal structure of the simulated outbreaks that are not apparent from goodness-of-fit measure alone.\\

{\noindent}The recurrence analysis provides a complementary characterization of the differences observed above by examining the organization and persistence of recurrent states in the reconstructed state space. Both the reported and calibrated stochastic time series data exhibited structured recurrence patterns, confirming that the outbreak dynamics were not characterized by random recurrence alone. At the same time, the RQA measures revealed a systematic distinction between the two representations. The calibrated stochastic simulations results consistently showed higher determinism, whereas the reported incidence exhibited higher laminarity and longer trapping times across all the three years. Thus, the calibrated model produces stronger diagonal recurrence organization, while the empirical outbreaks retain greater vertical recurrence and longer residence within the recurrent states. The temporal evolution of the RQA measures further showed that these differences vary throughout the outbreaks rather than remaining constant in time.\\

{\noindent}The CH and RQA analysis results provide a complementary evidence concerning the residual difference between the reported and calibrated model simulated time series data. The CH analysis indicates broadly similar but non-identical distributions of entropy and statistical complexity, whereas, the RQA analysis identifies differences in the organization and persistence of recurrent states. The agreement between these two analyses do not guarantee that they produce identical measures, but both reveal aspects of temporal organization that remain distinguishable after incidence-based calibration. This demonstrates that a model can reproduce the overall outbreak trajectory while retaining measurable differences in higher order dynamical properties.\\

{\noindent}In a broader sense, the current study demonstrates that fluctuations, seasonally driven time dependent transmission parameters and nonlinear dynamical analyses provide a strong complement to the conventional dengue outbreak model calibration. However, the statistical goodness-of-fit measures evaluate the degree of agreement between the simulated and observed incidence trajectories, complexity-entropy and recurrence analyses asses whether the underlying temporal organization of the outbreak dynamics is properly reproduced. The proposed framework, therefore, extends the model validation beyond incidence fitting and provides a more comprehensive approach for evaluating the dynamical consistency of stochastic dengue outbreak model. Although we demonstrated here using the dengue outbreaks data, the same framework can be applied to other stochastic models of other infectious transmission epidemic/pandemic models in which temporal dynamics play an important role.\\

{\noindent}We would like to acknowledge some limitations of this study. The analysis is based on laboratory-confirmed dengue incidence report from a single district over a three-year period. In addition, the stochastic SEIR--SEI model uses a simplified representation of dengue transmission and does not explicitly account for other environmental factors, spatial heterogeneity, human mobility, vector control interventions, vaccinations or reporting uncertainty. These factors may have an impact on both the observed incidence and the nonlinear dynamical properties of epidemic time series data. Future work may be the expansion of the current framework by incorporating other additional epidemiological and environmental parameters and processes into the proposed stochastic model and evaluating its applicability regions and longer observational records.

\section{Conclusion}
\label{sec:conc}
{\noindent}We develop a calibrated stochastic SEIR-SEI model to study the dynamics of dengue outbreaks in Imphal West district, Manipur, India, during the 2023-2025 transmission season. The incorporation of seasonal modulated transmission together with gradual activation mechanism enabled the model to reproduce the observed outbreak trajectories consistently across all three years. The sensitivity analysis further identified the transmission activation time $T_{50}$ as the dominant transmission-modulation parameter influencing model performance, emphasizing the importance of accurately representing the timing of seasonal transmission.\\

{\noindent}One of the important contributions of our study lies in extending the proposed model evaluation beyond the conventional incidence-based calibration. By integrating complexity-entropy and recurrence quantification analyses with stochastic dengue outbreak modeling, the proposed framework evaluates whether the calibrated model reproduces not only the reported outbreak trajectories but also the nonlinear dynamical organization of the observed incidence time series data. The results demonstrate that the calibrated stochastic model preserves the main dynamical characteristics of the reported dengue outbreaks while retaining measurable quantitative differences in higher-order nonlinear measures.\\

{\noindent}The analytical findings using various techniques mentioned above demonstrate that the good agreement at the trajectory level does not necessarily imply identical in higher order dynamical organization. Instead, nonlinear diagnostics reveal aspects of the outbreak dynamics that remain distinguishable after conventional calibration. The combined use of complexity-entropy and recurrence quantification analysis, therefore, provides complementary information for assessing the consistency of stochastic epidemic models.\\

{\noindent}Finally, this work demonstrates that nonlinear dynamical tools can provide valuable complementary information for evaluating epidemic models beyond traditional goodness-of-fit measures. The proposed framework may also be extended to incorporate additional epidemiological and environmental factors, offering further opportunities to investigate the dynamical organization of dengue transmission and other vector-borne infectious diseases.

\begin{acknowledgments}
T.S.S. would like to acknowledge the financial support provided by Manipur University. M.K.S. is partly supported by the National Fellowship for Scheduled Castes Students (NFSC), provided by the National Scheduled Castes Finance and Development Corporation (NSFDC) and the Department of Social Justice \& Empowerment, Ministry of Social Justice \& Empowerment, Government of India, with reference number 201610028460. M.S.S. would like to acknowledge Manipur University for providing the Seed Money Grant for Research and Development under the Research Promotion \& Incentive Policy.\\
\end{acknowledgments}

\section*{Author Declarations}
\subsection*{Conflict of Interest}
The authors have no conflicts to disclose.

\subsection*{Author Contributions}
\textbf{T.S.S.}: Conceptualization (supporting); Methodology (lead); Data curation (lead); Formal analysis (lead); Software (lead); Validation (equal); Visualization (equal); Writing – original draft (equal). \textbf{M.K.S.}: Conceptualization (lead); Methodology (supporting); Data curation (supporting); Formal analysis (supporting); Software (supporting); Validation (equal); Visualization (equal); Writing – original draft (equal).\textbf{R.K.B.S.}: Conceptualization (supporting); Supervision (equal); Writing – review and editing (equal). \textbf{M.S.S.}: Conceptualization (supporting); Supervision (equal); Writing – review and editing (equal). 

\section*{Data Availability}
Data available on request from the authors.

\end{document}